# Re-analysis of the Cassini RPWS/LP data in Titan's ionosphere. Part I: detection of several electron populations

**A. Chatain[1,2], J.-E. Wahlund[3], O. Shebanits[3,4], L. Z. Hadid[2,3], M. Morooka[3], N. J. T. Edberg[3], O. Guaitella[2], and N. Carrasco[1]**

[1] Université Paris-Saclay, UVSQ, CNRS, LATMOS, Guyancourt, France.

[2] LPP, Ecole polytechnique, Sorbonne Université, Institut Polytechnique de Paris, CNRS, Palaiseau, France.

[3] Swedish Institute of Space Physics, Uppsala, Sweden.

[4] Imperial College London, United-Kingdom.

Corresponding author: Audrey Chatain (audrey.chatain@ens-paris-saclay.fr)

**Key Points:**

- The Cassini Langmuir probe dataset in Titan's ionosphere is re-analyzed with a specific interest on the electron density and temperature.

- 2 to 4 cold electron populations with distinct potentials are observed.

- Electron populations vary with altitude and solar illumination suggesting origins linked to solar photons, magnetospheric particles and dust

**Abstract**

Current models of Titan's ionosphere have difficulties in explaining the observed electron density and/or temperature. In order to get new insights, we re-analyzed the data taken in the ionosphere of Titan by the Cassini Langmuir probe (LP), part of the Radio and Plasma Wave Science (RPWS) instrument. This is the first of two papers that present the new analysis method (current paper) and statistics on the whole dataset. We suggest that between 2 and 4 electron populations are necessary to fit the data. Each population is defined by a potential, an electron density and an electron temperature and is easily visualized by a distinct peak in the second derivative of the electron current, which is physically related to the electron energy distribution function (Druyvesteyn method). The detected populations vary with solar illumination and altitude. We suggest that the 4 electron populations are due to photo-ionization, magnetospheric particles, dusty plasma and electron emission from the probe boom, respectively.

## 1 Introduction

The Cassini mission explored the Saturnian system from 2004 to 2017 and gave us unprecedented insights on the ionosphere of Titan, the biggest moon of Saturn (Coates et al., 2007; Wahlund et al., 2005; Waite et al., 2005). This ionized environment hosts a complex ion chemistry that leads to the formation of organic aerosols (Vuitton et al., 2019; Waite et al., 2007). The Cassini mission investigated Titan at the occasion of 126 close flybys, and probed the ionosphere below the altitude of 1200 km for 57 of these flybys. There is a substantial coverage of measurements taken at different solar zenith angles, latitudes and seasons, enabling statistical analyses.

The study here is focused on electrons in Titan's ionosphere. The electron density was previously observed to be mainly governed by solar photons during the day (Ågren et al., 2009). Photoionization of the neutral molecules lead to a maximum electron density (~2000 cm$^{-3}$) around 1100 km altitude. The electron density have also been observed to vary (up to factor 2 at the peak) with the solar cycle (Edberg et al., 2013). On the nightside, electron transport from the dayside and the thermalization of magnetospheric electrons give a constant electron density of ~400-700 cm$^{-3}$ from 1200 km to < 950 km (Ågren et al., 2009; Cravens et al., 2009; Cui et al., 2009; Shebanits et al., 2017b).

Below 1100 km of altitude, positive, negative ions and dust have increasing densities, and start to affect the electrons (Shebanits et al., 2013, 2016; Waite et al., 2007). In particular, dust particles in a dense ionized environment tend to attract electrons and charge negatively (e.g. Farrell et al., 2009; Shukla & Mamun, 2015). It explains why the electron density decreases below 1100 km, while the density of ions and negatively charged dust increase and become the dominant negative charge carriers (e.g. Shebanits et al., 2013, 2016).

The temperature of the bulk electrons can be deduced from measurements by the Langmuir probe on-board Cassini. Ågren et al. (2009) and Edberg et al. (2010) derived the electron temperature profiles (assuming a single Maxwellian electron distribution) from 52 flybys down to ~900 km of altitude and stated a rather stabilized temperature below 1100 km. The observed values ranged from 0.03 to 0.06 eV ($\approx$ 350-700 K).

Several models have been developed to understand the processes at work in Titan's ionosphere (e.g. Galand et al., 2014 and references therein). Nevertheless, model results either predict too cold electrons (< 0.02 eV; < 200 K) at 900 km (Mukundan & Bhardwaj, 2018; Richard et al., 2011) or an overestimated electron density by a factor of 2 (Vigren et al., 2013, 2016). The addition of the negative ions to a photoionization model has been shown to decrease the discrepancy with the measurements (Shebanits et al., 2017a).

The present work aims to closely investigate the measurements of the electron density and temperature in the ionosphere of Titan, focusing on the effects of solar irradiation and altitude. The objective is to find clues of processes that are missing in the models. For this purpose, we re-analyzed the measurements acquired by the Langmuir probe (LP) part of the Radio and Plasma Wave Science (RPWS) investigation on-board Cassini, during the 13 years of the mission. This work is presented in two parts. In a first paper (this one), we detail the method used for the re-analysis of the data and the detection of several electron populations. A second paper (Chatain et al., 2021), referred as 'paper II', presents the results obtained for 57 flybys and discusses the origins of the detected electron populations.

## 2 Methods

### 2.1 Langmuir Probe measurements

The Langmuir probe on-board Cassini was built by the Swedish Institute of Space Physics (IRF). It was positioned below the radio antenna, on a 1.5 m boom to minimize electric perturbation from the spacecraft (Gurnett et al., 2004). The probe was a sphere of titanium of 5 cm in diameter, coated with titanium nitride, chosen foremost for its electrical work function stability, chemical inertness, high reflectivity and limited photoelectric effects. It is also very hard, enhancing solid particle impact resistance and eliminating hysteresis effects that are observed with other surface coatings (Wahlström et al., 1992). A small part of the stick (the stub) closest to the sphere was kept at the probe potential to ensure a symmetric potential pattern and sampling around a swept probe. A cleaning mode enabled removal of possible contamination on the surface. In this mode, the potential of the sphere was set at +32 V to induce sputtering of the probe with high energy electrons.

In this study, we used the measurements from the sweep mode of the probe. The current collected by the probe during the voltage sweep enables to deduce the electron density and temperature (among other useful derived parameters of interest). On Titan, the probe usually acquired double-sweeps, in 512 x2 voltage steps, from +4 V to -4 V, and back to +4 V, to give an idea of possible surface contamination effects. In addition, to detect any capacitive effects, for both of the sweeps the current was measured twice at an interval of 1 ms at each voltage step (see Figure A1 in the Appendix).

In this work, we first investigated the reproducibility of the measurements with the two voltage sweeps (named 'down' and 'up'), and the double measurements at each voltage step (resp. 'down1'-'down2' and 'up1'-'up2'). We observed that 'down1' and 'down2' (respectively 'up1' and 'up2') are always very close to each other (inferior to 1% difference): the current stabilizes quickly after each voltage step. We do not observe a strong difference between 'down' and 'up'. Further details are given in Appendix A. Besides, the fitting done on 'down' and 'up' sweeps give the same results for the electron densities and temperatures (see example in Figure S1 in Supporting Information).

In the following, we generally used the average of the decreasing part of the sweep ('down'). Nevertheless, we used also the average of 'up1' and 'up2' in the few cases where the decreasing part of the sweep was not acquired.

### 2.2 Theory to fit the electron and positive ion current

The Orbital Motion Limited (OML) theory is used to fit the voltage sweeps measured by the Langmuir probe. A correction is added in the attractive bias voltage region using the Sheath Limited theory (SL).

The OML theory, first described by Mott-Smith & Langmuir (1926), deduces information on the charged particles from the current measured by the probe, based on the conservation of energy and angular momentum. No particle is assumed to come from the probe itself and the bulk particle speed distribution is assumed Maxwellian. The OML theory is valid when the probe radius is smaller than the Debye length in the probed region. In these conditions, the electron and ion currents (respectively $I_e$ and $I_i$) are described by two different equations depending on the sign of $\Delta U = U_{bias} - U_P$ (Wahlund et al., 2009a; Whipple, 1965), $U_{bias}$ being the potential of the probe and $U_P$ the characteristic potential of the electron/ion population in the plasma. The electron current $I_e$ (or the ion current $I_i$) is expressed as a function of $I_{e,0}$ (respectively $I_{i,0}$) and $\chi_e$ (respectively $\chi_i$):

$$I_x = I_{x,0}\,(1 - \chi_x) \qquad \begin{array}{l} x = e \text{ if } \Delta U > 0, \, x = i \text{ if } \Delta U < 0 \\ \text{['attractive' part for electrons]} \end{array} \qquad (1)$$

$$I_x = I_{x,0}\,\exp(-\chi_x) \qquad \begin{array}{l} x = e \text{ if } \Delta U < 0, \, x = i \text{ if } \Delta U > 0 \\ \text{['repelling' part for electrons]} \end{array} \qquad (2)$$

The two equations join in $\Delta U = 0$. Then, the collected current $I_{x,0}$ is formulated as follows. It combines the effect of flow kinetic energy and thermal energy (Fahleson et al., 1974):

$$I_{x,0} = -A_{LP}.\,n_x.\,q_x.\,\sqrt{\frac{v_x^2}{16} + \frac{k_B T_x}{2\pi.\,m_x}} \qquad (3)$$

$A_{LP}$ is the surface area of the probe. $n_x$, $q_x$, $v_x$, $T_x$ and $m_x$ are respectively the density, charge, velocity, temperature and mass of electrons or ions, and $k_B$ is the Boltzmann constant.

The parameter $\chi_x$ depends on $\Delta U$ and is expressed as:

$$\chi_x = \frac{q_x.\,\Delta U}{\frac{m_x v_x^2}{2} + k_B T_x} \qquad (4)$$

In the case of electrons, the flow kinetic energy term can be neglected compared to the thermal term. It is the opposite in the case of ions (positive and negative), which are heavier than electrons and transported along the ion ram flux. When $\Delta U \ll 0$, the collected current is governed by positive ions while it is dominated by electrons when $\Delta U > 0$.

In Titan's ionosphere, the Debye length ($\lambda_D \approx 2.5 - 10.5\ cm$) is similar to the radius of the probe ($r_p = 2.5\ cm$). It is the limit of validity to use the OML theory. Therefore, a small correction is added by using the Sheath Limited (SL) theory (Bettinger & Walker, 1965; Whipple, 1965), valid in a Maxwellian plasma. It adds the dependence of the sheath thickness ($s$) with the potential and size of the probe and changes the expression of the attractive part. For electrons, it gives for $\Delta U > 0$:

$$I_e = I_{e,0} \times \left[ 1 + \xi \times \left( 1 - \exp\left( +\frac{\chi_e}{\xi} \right) \right) \right]$$

$$with\quad \xi = \left( \frac{s}{\rho} + 1 \right)^2 - 1, \qquad \rho = \frac{r_p}{\lambda_D} \tag{5}$$

$$and\quad s = 0.83 . \sqrt{|\chi_e|} . \rho^{\frac{1}{3}}$$

The expression of the sheath thickness ($s$) is empirically obtained by Bettinger and Walker (1965) in the case of spherical bodies. These expressions converge to the OML expressions when $s \gg r_p$.

The electron temperature and the Electron Energy Distribution Function (EEDF) can be obtained from the 'transition region', the part of the voltage sweep where the electron current starts to dominate the positive ion current, i.e. for $\Delta U$ being slightly negative (of a few tenths of volts in the conditions of Titan's ionosphere).

In this transition region, the positive ion current is not always negligible compared to the electron current. To obtain the electron current, the measured (total) current is therefore corrected from the positive ion part, fitted by a linear curve (see Equation (1)). On the negative bias voltage zone, a detailed study of the ion current can give information on the ions properties (Shebanits et al., 2013). Nevertheless, in the transition region, the uncertainty on the ion current fit is negligible compared to the total current error bars (see Appendix C.1). This is because the ion current is small compared to the electron current and it linearly varies with an increasing potential. Therefore, it does not affect the large and quickly varying electron signatures. An example showing the total current, the ion current fit, the electron current and their error bars is given in Supporting Information Figure S2.

### 2.3 Example of electron current collected in Titan's ionosphere: observation of several electron populations

The electron part is fitted with Equations (2) and (5). Note that we always fit the curve I(U$_{bias}$), not its derivatives. Nevertheless, using only one electron 'population' (i.e. one combination of Equations (2) and (5)) results in a poor fit of the electron current collected in the ionosphere. Generally, a second population has to be added. It leads to two different sets of ($U_P$, $n_e$, $T_e$). An example is shown in Figure 1. This method is usually used in Langmuir probe sweep data analysis (Ågren et al., 2009; Edberg et al., 2011; Wahlund et al., 2009a, 2009b). This necessity to fit 2 populations is discussed further in Section 3.2.1.2.

Below 1300 km, two maxwellian populations of electrons are often still not sufficient to correctly fit the electron current. A third, and in rare occasions a fourth, population is required. An example with three populations is shown in Figure 2 and examples with one and four populations are given in Supporting Information Figure S3 and S4. The necessity to use 3 or 4 electron populations for the fit is discussed in Sections 3.2.2.2 and 3.2.3.2. In conclusion, the analysis of voltage sweeps in the ionosphere of Titan often leads to 2-4 sets of $(U_P, n_e, T_e)$ per sweep. The fitting procedure is detailed in Appendix B.

Nevertheless, the use of several Maxwellian populations is not physically correct because if all electrons were thermalized, we would observe only one Maxwellian population. This is an approximation. Among the four populations, one is certainly thermalized, but the three others are not yet fully thermalized. The electron thermalization is given by the energy equation for an ionospheric plasma. The thermalization time constant depends on several loss ($L_e$) and source ($Q_e$) terms, including rotation and vibrational collisional transitions, but basically one could write (neglecting conduction and transport, see e.g. Galand et al., 2014):

$$\frac{3}{2} \, n_e \, k_B \, \frac{\partial T_e}{\partial t} \approx \sum Q_e - \sum L_e \qquad (6)$$

An approximate characteristic thermalization time constant can therefore be calculated from $\Delta t \sim 3/2 \, n_e \, k_B \, T_e \, / \, L_{tot}$ in Titan's ionosphere. $10^{-3}$-$10^1$ eV/cm$^{-3}$/s are typical values for the total loss ($L_{tot}$) below 2000 km, giving $\Delta t > 100$ s for 1 eV electrons. Thus, thermalization of electrons is a reasonably slow process and significant deviations from a Maxwellian distribution could exist from this rough analysis. Therefore, the non fully thermalized populations should in theory be fitted with other distribution functions, like Kappa distributions (Pierrard & Lazar, 2010). However, all of these populations are cold ($T_e < 0.1$ eV). They are nearly thermalized, and therefore the corresponding Kappa distributions are very similar to Maxwellians. Besides, a Kappa distribution requires more fitting parameters than a Maxwellian. A precise fit consequently requires high resolution data, which is not adapted to the low resolution Cassini Langmuir probe data. Because all of these reasons, we made the approximation to use Maxwellian distributions for all the four populations.

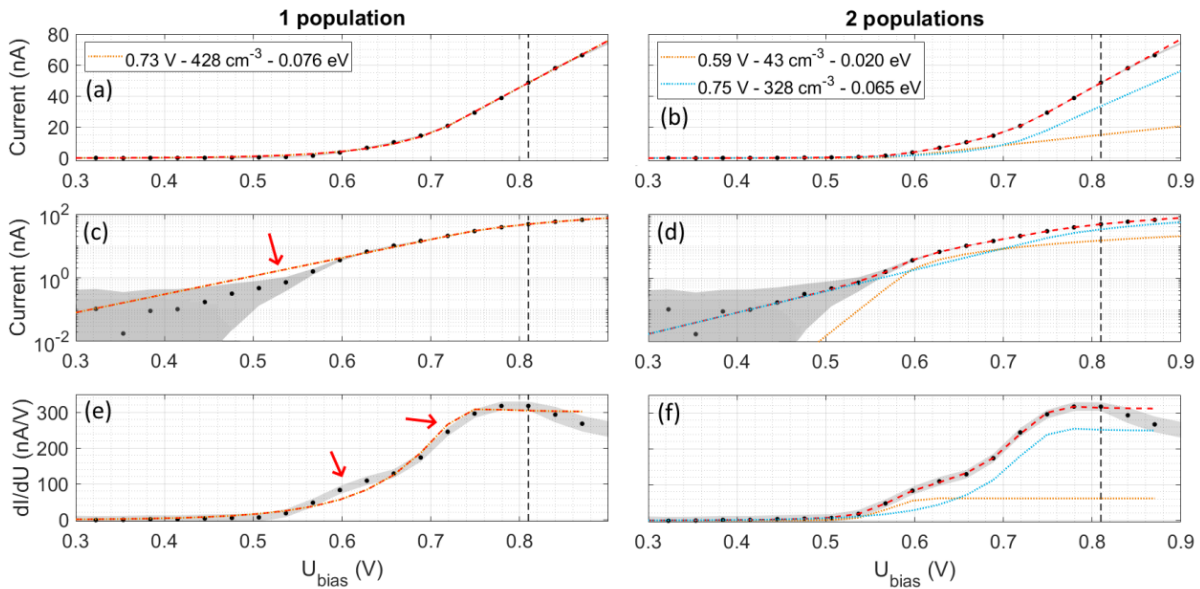

**Figure 1**. Fitting of the voltage sweep acquired during T50 at 1241 km altitude. In (a,c,e), one electron population is used for fitting. In (b,d,f), two electron populations are used for fitting. The fit results $(U_p, n_e, T_e)$ for each population are indicated on the graphs. The same current data points and fit are presented with three different plots: in linear scale (a and b), in logarithmic scale (c and d) and by their first derivative (e and f). Grey zones correspond to the data error bars (their computation is described in Appendix C.1). Red arrows indicate poorly fitted zones. The part of the curves after the dashed line corresponds to a part possibly distorted by the effect of a logarithmic pre-amplifier, and should not be used without further consideration.

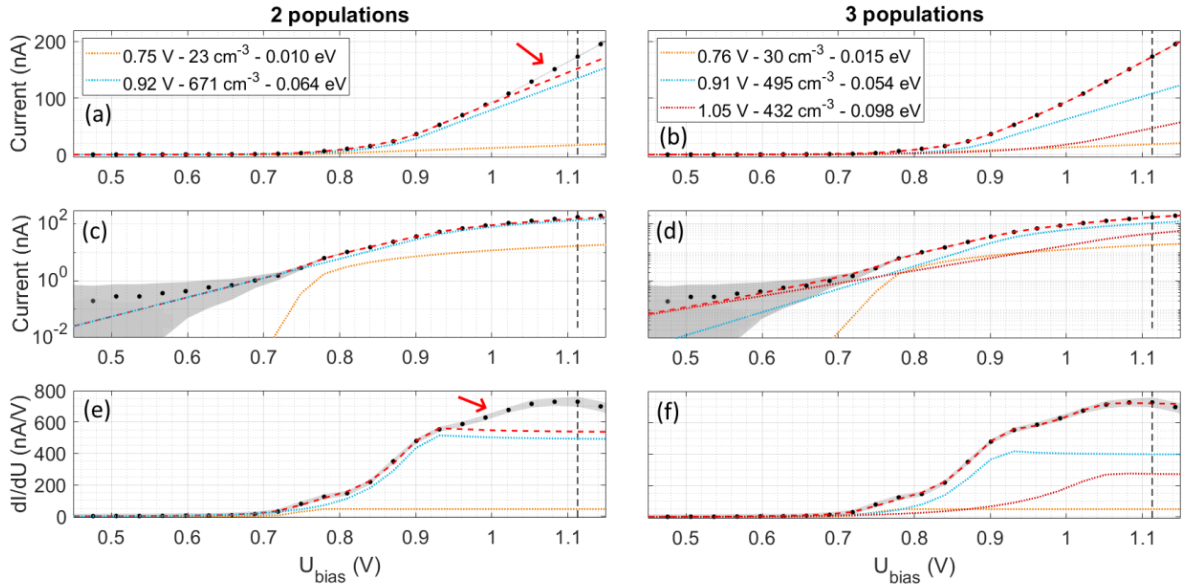

**Figure 2**. Similar to Figure 1, but at 1125 km altitude. (a,c,e) Two electron populations are used for fitting. (b,d,f) Three electron populations are used for fitting.

### 2.4 The second derivative of the electron current (d²I/dU²): an effective method to detect the electron populations

The second derivative of the current is a good indicator of the presence of several electron populations. Each peak corresponds to a population, and its area is globally proportional to the electron density associated. Appendix D presents the link between the second derivative of the electron current and the Electron Energy Distribution Function (EEDF). Figure 3 shows d²I/dU² associated with the fitted sweeps plotted in Figure 1 and 2. Cases with 1 and 4 populations are shown in Supporting Information Figure S5.

The data points are at very low resolution, with only four to ten points per peak in d²I/dU². Consequently, the computation of the second derivative is strongly impacted. This resolution effect is investigated on the fit curves and illustrated in Figure 3. From the equations of the fit curves (Equations (2) and (5)), we created two sets of artificial data I(U): one with a lot of data points (> 200 points per peak, named 'high resolution'), and one with as many data points as the real data (4-10 points per peak, named 'low resolution'). We computed the second derivative of these two artificial datasets ('low resolution' case in red line, 'high resolution' case in orange dotted-dashed line), and compared them to the d²I/dU² computed from the data (black dots).

The confidence intervals on the fit results ($U_P$, $n_e$, $T_e$) for all the populations are also computed (see details in Appendix C.2) and are plotted in the Figures of section 3.2. The uncertainty is larger in the cases with more electron populations, in particular when their parameters $U_P$ are close.

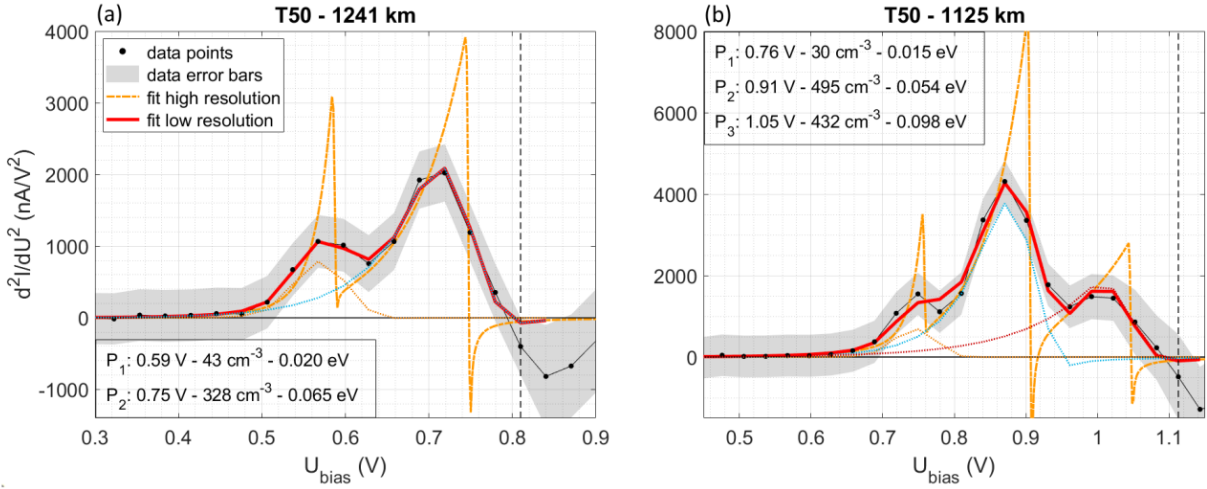

**Figure 3.** Second derivative of current for the sweeps shown in (a) Figures 1 and (b) Figure 2. The fitted curves shown in Figures 1 and 2 are derived to obtain d²I/dU², at the resolution of the data points (red line) and at higher resolution (orange dotted-dashed line). The fit results ($U_p$, $n_e$, $T_e$) for each population are indicated on the graphs. Grey zones correspond to the data error bars (their computation is described in Appendix C.1). The part of the curves after the dashed line corresponds to a part possibly distorted by the effect of a logarithmic pre-amplifier, and should not be used without further consideration.

## 3 Results and discussions

### 3.1 Variation of the electron populations with altitude and Solar Zenith Angle (SZA)

#### *3.1.1 Variation with altitude*

Voltage sweeps evolve during a flyby. The number of observed peaks in the second derivative of the current (d²I/dU²) varies with altitude. Therefore, the number of electron populations (and consequently the electron distribution) changes with altitude for a given flyby. We focus on the part of the ionosphere below 1200 km. The Langmuir probe took measurements at these altitudes at the occasion of 57 flybys.

The curves of d²I/dU² as a function of the altitude obtained from the 57 flybys allow to easily classify Titan flybys in three groups. Each group has a different number of electron populations, with different characteristics. They are described in Table 1 and examples of profiles for the three cases are shown in Figure 4. All the profiles have similar d²I/dU² curves around 1250-1200 km, showing usually three peaks. However, this curve evolves differently with altitude in the three cases. At lower altitude, in all cases, there is a main hump formed either by one large peak, two or three peaks (further named $P_1$, $P_2$, $P_3$ for clarity). In addition, for the profiles of the group G3, a small hump detached from the main one (P$_4$), at higher voltage, is observed below ~1200 – 1150 km altitude.

**Table 1.** Definition of the 3 flyby groups and their corresponding flybys. The flybys studied in details in Figure 4 and Section 3.2 are indicated in bold.

| Name of the flyby group | Characteristics of d²I/dU² at lower altitude | Flybys (T#) |
|---|---|---|
| G1 | 2 separated peaks | 21, 25-28, 55, **118**-119 |
| G2 | 3 peaks | 5, 16, 46, 50, 56-**59**, 65, 117, 121 |
| G3 | 1 large peak (or 2 close) + 1 further, at higher U$_{bias}$ | 17, 20, 23, 39-43, 47-48, 51, 83-88, 91-92, 95, 100, **104**, 106-108, 113, 126 |
| G1/G2 | with different characteristics on inbound and outbound | 29-30, 32, 120 |
| G2/G3 | | 18-19, 36, 49, 61, 70-71 |

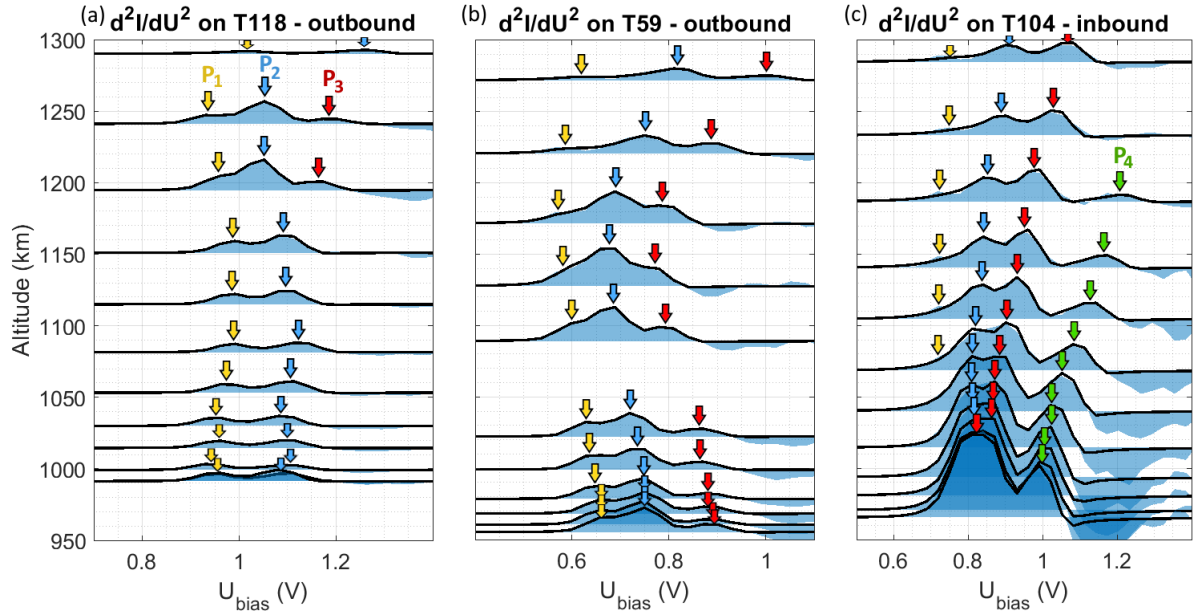

**Figure 4.** Examples of d²I/dU² profiles with altitude: (a) T118 (from G1), (b) T59 (G2) and (c) T104 (G3). The second derivative of the electron current from Langmuir probe data (blue area) is superimposed to d²I/dU² from the fit (black line). The scaling of d²I/dU² is the same for all the plots. Arrows indicate the peaks corresponding to the 4 populations.

### 3.1.2 Variation with Solar Zenith Angle (SZA)

The detection of populations P$_3$ and P$_4$ are strongly dependent on the Solar Zenith Angle (SZA). Figure 5 shows the repartition of the electron populations with altitude and SZA in the case of 57 Cassini flybys. The population P$_4$ appears always below ~1200 km and at SZA < 80-90° (such flybys belong to the group G3), and the cases without a $P_3$ population at lower altitude (group G1), are observed only on the nightside, at high SZA (>~120°).

The distribution of the 57 flybys in three groups is then closely correlated to their SZA. The group G1 corresponds to flybys on the far nightside (SZA >~120°), G2 to flybys on nightside close to the terminator, and G3 to dayside (SZA < 80-90°).

In paper II, the densities of $P_3$ and $P_4$ electron populations are found to be correlated to extreme UV fluxes and (positive and negative) ion densities. This suggests that these populations are linked to the photo-ionization processes occurring in Titan's ionosphere. Suggestions on the electron population origins are further discussed in Sections 3.2.2.3 and 3.2.3.3.

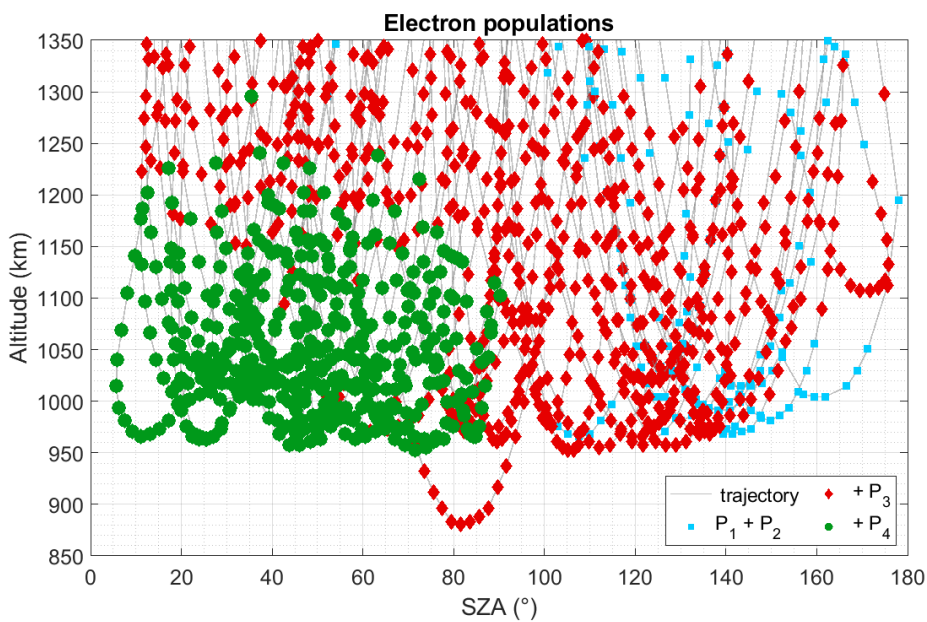

**Figure 5.** Detection of electron populations as a function of altitude and SZA, from 57 flybys. Blue squares represent sweeps with only populations $P_1$ and $P_2$, red diamonds show the presence of $P_3$, and green dots are for $P_4$.

In conclusion, the variation with altitude and SZA of the detected peaks of $d^2I/dU^2$ can be schematized as in Figure 6. Each peak is attributed to an electron population among $P_1$, $P_2$, $P_3$ and $P_4$. At ~1200-1300 km, $d^2I/dU^2$ graphs are very similar at all SZA, with three populations ($P_1$, $P_2$, $P_3$). Below 1200 km, $d^2I/dU^2$ varies differently depending on SZA: (G1) on the far nightside (SZA >~120°), $P_3$ disappears; (G2) close to the terminator, the three populations are kept at lower altitudes, with an increase of the area of the $P_2$ peak; (G3) and on dayside, $P_2$ and $P_3$ grow, $P_1$ becomes negligible and $P_4$ appears. In some occasions, the $P_3$ peak strongly grows and covers $P_2$.

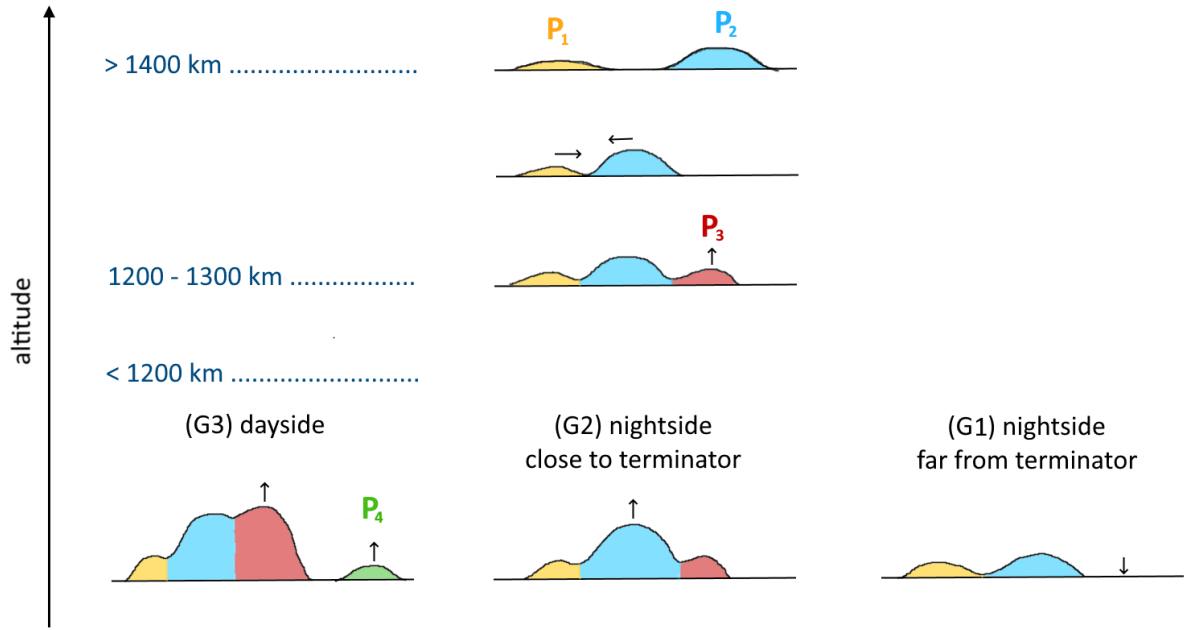

**Figure 6.** Summary scheme of the evolution of d²I/dU² peaks with altitude in the 3 cases: nightside far from terminator (G1), nightside close to terminator (G2) and dayside (G3). Small arrows indicate the evolution trends with decreasing altitude.

### 3.2 Electron densities and temperatures

The results on electron density and temperature are discussed separately for each of the 3 groups (G1, G2 and G3), in the following sub-sections (respectively 3.2.1, 3.2.2 and 3.2.3).

#### 3.2.1 Nightside far from the terminator (group G1)

##### 3.2.1.1 Typical altitude profiles

Figure 7 shows an example of inbound (solid lines) and outbound (dashed lines) profiles of the electron density and temperature typical of the far nightside (G1). Following comments are also valid for the other G1 flybys. The confidence interval at 95% is indicated with altitude (see computation in Appendix C). The dominant population observed is $P_2$ (blue), with a rather constant temperature around 40-60 meV (460-700 K) below 1150 km altitude (and in some cases also below 1250 km). The density of this population peaks at an altitude varying between 1150 and 1250 km altitude. The population with the lowest $U_p$ ($P_1$, orange) is very constant with altitude. Its density is low, less than 100 cm$^{-3}$. Its temperature is around 10-20 meV (120-230 K), at the detection limit of the probe. In addition, a third population appears ($P_3$, red) at the ionospheric peak, when the total electron density exceeds 500 cm$^{-3}$. It has a temperature higher than the other populations, around 60-90 meV (700-1040 K). Its presence is discussed in Section 3.2.2.3.

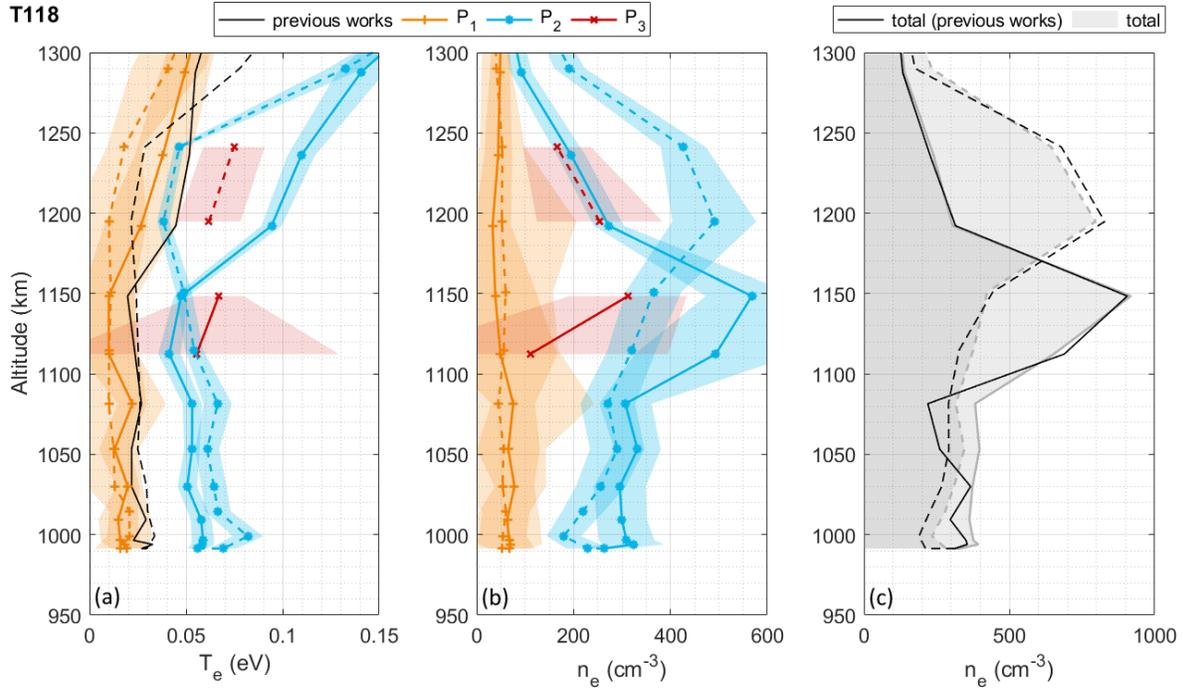

**Figure 7.** Electron temperature (a) and density (b,c) profiles in the case of T118, an example of a G1 group flyby (far nightside); results for the different populations P₁, P₂ and P₃ (colors). All the solid curves are inbound and dashed are outbound. The fit confidence intervals at 95% are indicated in colored shaded area. The total density (sum of the densities of all the populations) is indicated in grey area. The electron temperature and density obtained in previous works (Ågren et al., 2009; Edberg et al., 2010, 2013; Garnier et al., 2009) are showed with black lines.

In all the analyzed flybys, the total electron density obtained is globally consistent with previous measurements (Ågren et al., 2009; Edberg et al., 2013; Garnier et al., 2009), within the given error bars. In the case of G1 flybys, the previous analysis usually gives an electron temperature between the ones found for P₁ and P₂, and therefore sometimes colder than the dominant population (P₂) temperature. For instance, in the case of T118 shown in Figure 7, they obtained a temperature not representative of the main population of electrons (P₂). These previous works have been done on the same dataset, but with a different method of analysis. The authors also used several Maxwellians to fit the data (up to 3), but the use of several small Maxwellians was only to help to fit as well as possible the signal with one 'main' Maxwellian (that was usually the second one). They deduced the electron temperature from the fitting result of this 'main' Maxwellian. Here we go one step further by identifying the Maxwellians into electron populations. This enables to have one temperature per population and consequently, deduce the temperature of the dominant electron population without ambiguity.

### 3.2.1.2 Necessity to fit multiple populations

Even if the population P₁ is not abundant, it is necessary to fit it separately from the main population P₂ to avoid a hot bias on the determination of the temperature of P₂. For comparison, we studied the case where the current curve is fitted using only one population, as shown in Figure 1a. The results are compared in Figure 8. Forcing only one electron population ($P_2$, dark blue) overestimates the temperature of the dominant population ($P_2$, light blue) by ~0.01 eV (~120 K). In addition, the total electron density increases of ~10%.

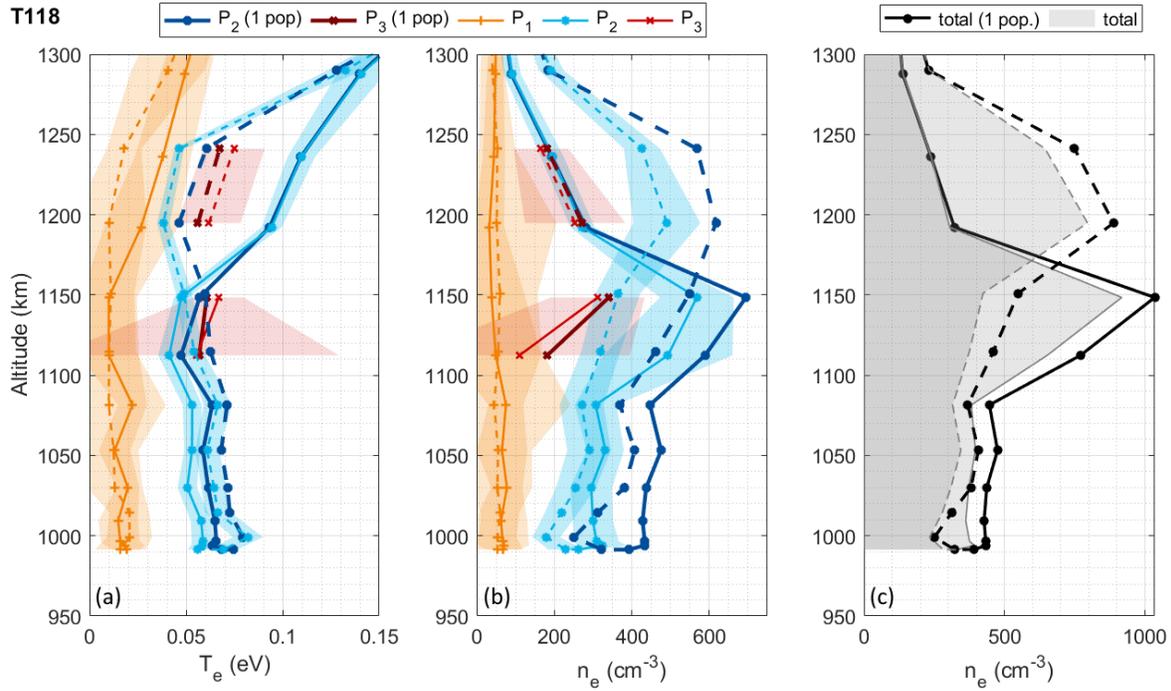

**Figure 8.** Same as Figure 7, superimposed to the $T_e$ and $n_e$ profiles obtained with only one population to fit $P_1$ and $P_2$ when their potentials are close enough (i.e. below 1200 km for inbound, and below 1300 km for outbound).

### 3.2.1.3 Discussion on the origin of $P_1$ and $P_2$

On the nightside far from the terminator, ionization is not governed by solar photons, but by energetic particles coming from the magnetosphere (Ågren et al., 2007; Cravens et al., 2008; Galand et al., 2014). We therefore suggest that the dominant electron population ($P_2$) is the bulk of thermalized electrons initially formed by processes involving particle precipitation.

The other population present at all altitudes, $P_1$, is certainly due to electrons emitted by the nearby probe boom, from collisions with energetic particles. Such electrons are created on the nearby surfaces, they have low energy and are easily caught back. Therefore, $P_1$ corresponds to the population with the lower potential $U_p$.

$P_3$ is discussed in the following sections, as it is observed in higher quantities in groups G2 and G3.

### 3.2.2 Nightside close to the terminator (group G2)

#### 3.2.2.1 Typical altitude profiles

Flybys on the nightside close to the terminator (~90° < SZA ~< 140°) show the presence of the third population at all altitudes (see Figure 9).

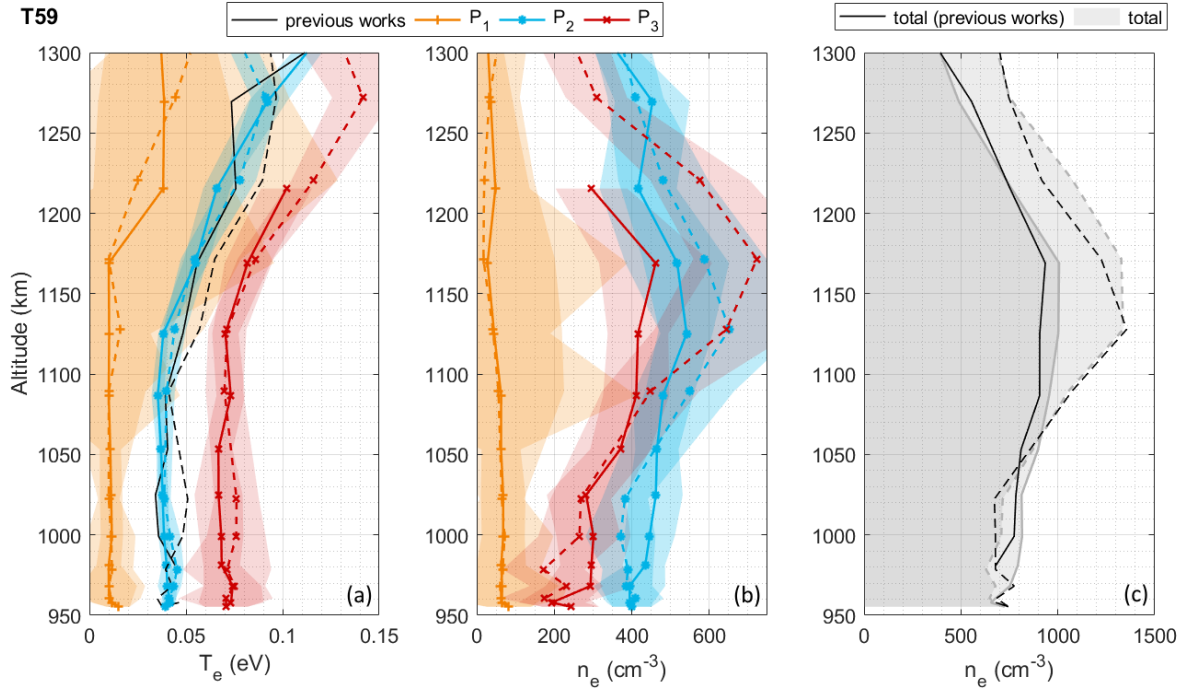

**Figure 9.** Same as Figure 7, in the case of T59, an example of a G2 group flyby (nightside close to the terminator).

The total electron density is globally higher than for G1 flybys (far nightside), which is expected because of the closer influence of photoionization on dayside (Ågren et al., 2009). The temperatures observed for the three populations are similar to the ones of G1 flybys: $P_1$ below 20 meV (~230 K, orange), $P_2$ generally between 30 and 50 meV (350-580 K, blue) and $P_3$ at 60-80 meV (700-930 K, red).

#### 3.2.2.2 Necessity to fit multiple populations

The addition of the third population ($P_3$) does not disturb the fitting of the two first populations. Nevertheless, its removal decreases strongly the total electron density. This effect is showed in Figure 10. It compares the results obtained in Figure 9 with results of a fit that does not include the population $P_3$, as plotted in Figure 2 (a). Then, the temperatures obtained for $P_1$ and $P_2$ are almost unchanged, but the total electron density decreases of 25%.

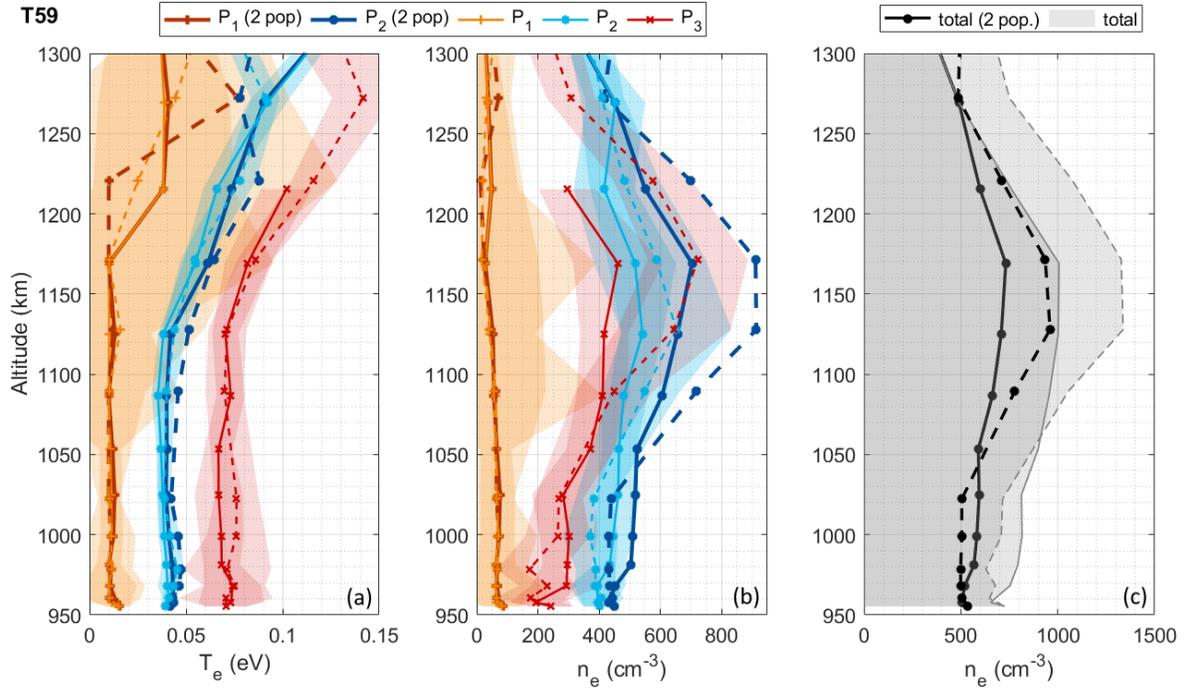

**Figure 10.** Same as Figure 9, superimposed to the $T_e$ and $n_e$ profiles obtained for $P_1$ and $P_2$ when $P_3$ is not fitted.

#### 3.2.2.3 Discussion on the origin of P₃

The fact that $P_3$ is observed on dayside (G3) but generally not on the far nightside (G1), suggests that it is due to the efficient photo-ionization processes happening on the dayside. Their influence is extended by transport to the nightside close to the terminator. Remnants are still observed at higher SZA at the ionospheric peak. Such transport of photoelectrons and their secondaries has already been suggested in previous works (Ågren et al., 2009; Cui et al., 2010; Galand et al., 2014; Shebanits et al., 2016). Nevertheless, models suggest that it is not the main electron production process in the nightside below 1200 km. This is consistent with our observations showing that it is not $P_3$ but $P_2$ which is the main electron population on the nightside (especially in cases far from the terminator). The advanced analysis done in paper II shows a strong correlation between $P_3$ electron density and the extreme UV flux. This also supports the photo-ionization hypothesis.

### 3.2.3 Dayside (group G3)

#### 3.2.3.1 Typical altitude profiles

On dayside, a new population ($P_4$) appears in the LP sweep data at a potential $U_P$ higher than for $P_1$, $P_2$ and $P_3$. Its temperature is generally higher than for the other populations, up to 0.1 eV (~1160 K, green). This population appears only at altitudes below ~1200-1150 km and its density increase with decreasing altitude. In some cases (T17, T20, T85, T88, T104, T108), it even becomes the dominant population below 1050-1000 km, as shown in Figure 11. Contrary to the constant profiles on the nightside, the temperatures of all the populations decrease slightly with altitude, of ~0.03 eV (~350 K) in 200 km.

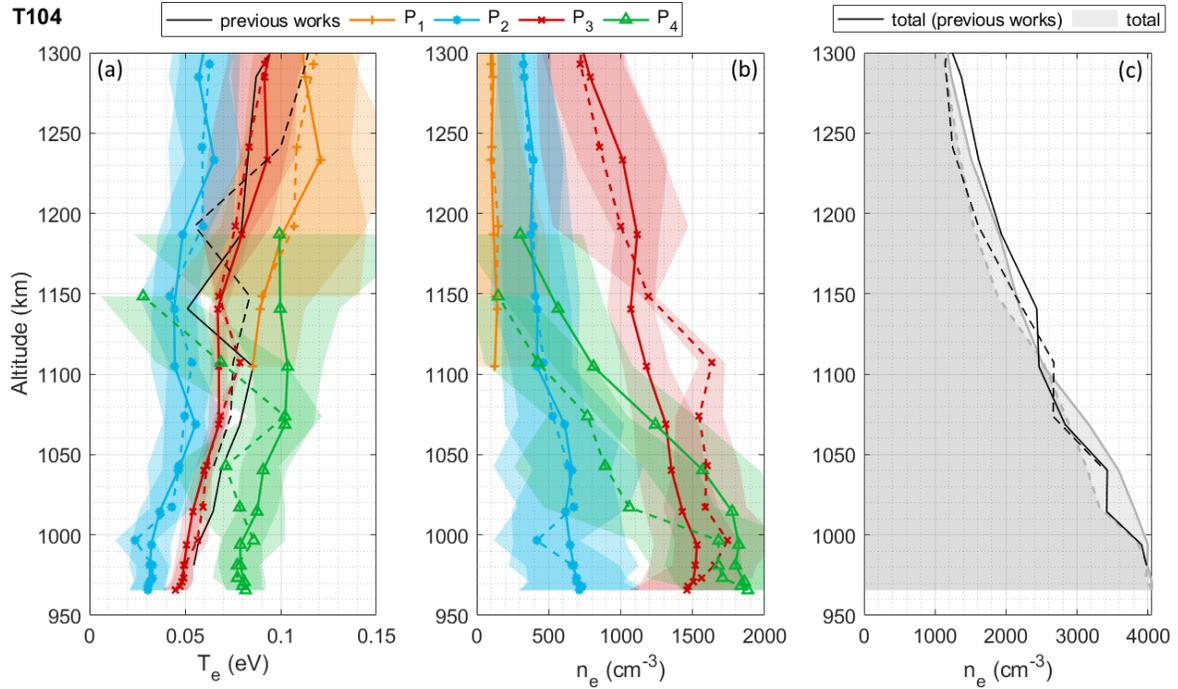

**Figure 11.** Same as Figure 7, in the case of a T104, an example of a G3 group flyby (dayside).

### 3.2.3.2 Necessity to fit $P_4$ and covering of $P_2$ hump by $P_3$

The electron population $P_4$ is often non-negligible and should always be taken into account when present. An example of the results obtained if $P_4$ is not fitted is given in Supporting Information Figure S7. The results are similar to Section 3.2.2.2: the ($U_p$, $n_e$, $T_e$) results for the other populations are not impacted, but the total electron density is underestimated by up to 30%.

In some cases, the fitting of $P_2$ and $P_3$ can be ambiguous, especially at lower altitudes. The use of two distinct populations is required to obtain a very good fitting (examples of sweeps are given in Supporting Information Figure S8) and is consistent across all altitudes (see Figure 4 for T104). Nevertheless, we investigated the effect on the results if the $P_2$ population were not fitted in the ambiguous cases. The result is that the covering of $P_2$ hump by $P_3$ at lower altitudes in G3 flybys happens smoothly: at one point, fitting with two or three populations gives globally the same results: $P_2$ becomes negligible compared to $P_3$. An example of the $n_e$ and $T_e$ profiles obtained when $P_2$ is not fitted at lower altitude is given in Supporting Information Figure S9. This does not affect our main results: the total electron density stays globally the same, the temperature of the main population is not impacted and the population $P_4$ generally stays in the range of its large error bars.

### 3.2.3.3 Discussion on the origin of P₁ and P₄

Similarly to the nightside, $P_1$ electrons on the dayside are certainly emitted by the probe boom. However, on dayside, the photoemission is also possible. In this case, the emitted electrons observed are hotter than on nightside. Such photoelectrons are commonly observed in Langmuir probe sweeps (Ågren et al., 2009; Gustafsson & Wahlund, 2010; Wahlund et al., 2009a; Wang et al., 2015).

The population $P_4$ is only detected on dayside. We argue here that it is not an instrumental effect but rather due to dusty plasma:

- Energetic photons and particles can detach electrons from the surface of the spacecraft. A recent study quantified the effect of the neaby spacecraft on Langmuir probe measurements applied to the case of the MAVEN mission on Mars (Ergun et al., 2021). Such a study has not yet been done on Cassini's Langmuir probe. Nevertheless, in the ionosphere of Titan the electrons are cold enough to prevent these spacecraft electrons to reach the Langmuir probe (Wang et al., 2015). Indeed, the Debye length in the ionosphere is always below 10.5 cm, far lower than the distance between the spacecraft and the Langmuir Probe (which is on a 1.5 m boom).

- The spacecraft motion cannot explain the presence of several electron populations. With energies from 0.02 to 0.08 eV in the ionosphere, the electrons of the various populations have a thermal velocity of 85 to 170 km/s, much larger than the spacecraft velocity (~6 km/s). Therefore, there should be no difference in the electron collection from all around the probe.

- The perturbation of the collected current by organic deposit on the surface of the probe is very unlikely. In theory, the complex organic chemistry happening in the ionosphere (Vuitton et al., 2019; Waite et al., 2005) could possibly lead to the formation of a deposit on the surface of the probe during its stay in the ionosphere, as can be observed in laboratory experiments mimicking Titan's ionospheric conditions (Gautier et al., 2012; Imanaka et al., 2012; Mahjoub et al., 2012). Nevertheless, the comparison to laboratory conditions suggests a very low coating speed. For example, experiments performed in the PAMPRE laboratory setup observe a coating of ~7 nm/min (Carrasco et al., 2016). Plasma conditions in the laboratory are different than in Titan's ionosphere. If we roughly consider that the coating efficiency is proportional to the ion density (that we suppose equal to the electron density here), then we can deduce the coating speed in Titan's ionosphere. Electron density is ~$10^3$ cm$^{-3}$ in Titan's ionosphere, and ~$10^9$ cm$^{-3}$ in the PAMPRE experiment (Chatain et al., 2020; Wattieaux et al., 2015). It roughly gives a coating speed of $10^{-5}$ nm/min on Titan. Cassini flybys in the ionosphere lasted ~15 min, and the probe was cleaned between two flybys (see Section 2.1). From this quick computation, we can conclude that organic coating on the probe is certainly negligible. Nevertheless, even if a coating formed on the probe during the flyby, it would not explain the apparition of populations $P_3$ and $P_4$. An organic film coating, depending on the surface coverage and dielectric properties, could possibly induce the detection of several electron components in the sweeps, but as the coating grows during the flyby, we then expect the artefacts to increase. However, this is not the case. Three points suggest that there is no organic coating on the probe: (1) Similar results are obtained in the inbound and outbound parts of the flyby when they are located at similar SZA (see Figure 11 as an example). (2) Over the 57 analyzed flybys, results are only dependent on the altitude and SZA, and there is no correlation to the history of the flyby trajectory (see Figure 5 and paper II). (3) No hysteresis is observed: down and up sweeps give the same results (see Section 2.1, Appendix A and Supporting Information Figure S1).

- $P_4$ is possibly due to a dusty plasma effect. Indeed, it is not predicted by current ionospheric models that do not include the effect of the charged dust grains or aerosols on the electron populations. The fourth electron population is only detected in the regions where dusty plasma has been previously measured (Shebanits et al., 2016). This suggests that the charged dust grains are at its origin, either as a source or a disruption of the ambient electrons. This possibility is investigated in paper II, where the results obtained on the 57 Cassini flybys in the ionosphere are analyzed in detail. In particular, they are compared to the extreme UV fluxes and the positive and negative ion densities computed by Shebanits et al. (2016 and 2017b). We observed a strong correlation with the $P_4$ electron density for the three parameters. As heavy negative ions are dust particles, we conclude that $P_4$ electrons are possibly formed from the exposure of organic dust to energetic solar photons. We suggest two mechanisms: direct photo-emission or heating followed by thermo-emission.

## 4 Conclusions

The re-analysis of the Langmuir probe data in Titan's ionosphere brought new insights on the behavior of the electrons in this environment.

The detailed analysis of the data revealed the presence of several electron populations, at different potentials, and with different electron densities and temperatures. We have shown that the second derivative of the electron current collected by the Langmuir probe during a voltage sweep is a useful tool to detected the presence of several electron populations, especially as it is linked to the electron energy distribution function (EEDF) (Druyvesteyn, 1930). With the evolution of the second derivative of the current with altitude we observed the evolution of the electron populations with altitude in three different conditions: far nightside (T118), nightside close to the terminator (T59) and dayside (T104).

Two populations (with the lower potentials) are present in nearly all conditions. The third population is not observed on the far nightside, and the fourth is detected on dayside below 1200-1150 km. We have investigated the fitting of the electron current curve with 2 to 4 populations with a systematic approach on all flybys that validated the presence of 2, 3 or 4 populations only depending on the altitude and the solar illumination. A rigorous testing on the number of fitted populations shows that the fitting of all the detected populations (i.e. the whole electron distribution) is necessary to obtain a correct fit and retrieve the total electron density.

Based on our observations, we suggested possible origins for the four populations:

- $P_1$ electrons, with the lowest potential and a low density, are emitted by the probe boom;
- $P_2$ electrons, found even on the far nightside, are possibly induced by particle precipitation;
- $P_3$ electrons, not present on the far nightside, are likely to be related to photo-ionization;
- and $P_4$ electrons, observed on dayside at lower altitudes (exactly where heavy negative ions and dust are formed) are plausibly linked to dusty plasma effects.

These four populations are studied in more details in paper II that performs a complete investigation on all the Cassini Langmuir probe dataset in the ionosphere of Titan. Based on these statistical results, the suggestions on the origins for the four populations are discussed further in paper II, and more detailed mechanisms are suggested. To go further in the investigation of the electron populations and get quantitative explanations of their origins, the next step will be to include these suggested processes in ionospheric models.

## Acknowledgments and Data

The Swedish National Space Board (SNSB) supports the RPWS/LP instrument on board Cassini. A.C. acknowledges ENS Paris-Saclay Doctoral Program. A.C. is grateful to Anders Eriksson, Jean-Pierre Lebreton, Erik Vigren, Ronan Modolo and the two reviewers for fruitful discussions on this project. O.S. acknowledges funding by the Royal Society grant RP\EA\180014. N.J.T.E. was funded by the Swedish National Space Board under contract 135/13 and by the Swedish Research Council under contract 621-2013-4191. N.C. acknowledges the financial support of the European Research Council (ERC Starting Grant PRIMCHEM, Grant agreement no. 636829).

All Cassini RPWS data are archived in the Planetary Data System (PDS) Planetary Plasma Interaction (PPI) node at https://pds-ppi.igpp.ucla.edu on a pre-arranged schedule.

## Appendix A. Measurement reproducibility

We investigated the reproducibility of the measurements with the two voltage sweeps (named 'down' and 'up'), and the double measurements at each voltage step (resp. 'down1'-'down2' and 'up1'-'up2'). Figure A1 illustrates the definition the data points 'down1', 'down2', 'up1' and 'up2' at a given potential. Figure A2 shows the comparison of sweeps obtained with these different data points.

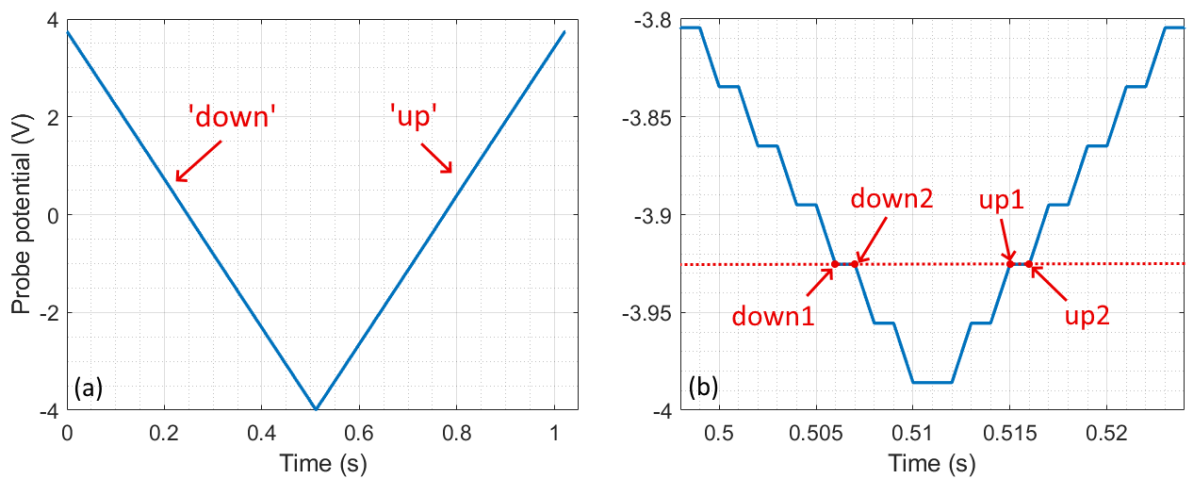

**Figure A1.** Variation of the probe potential with time during a double sweep measurement. (a) On the total duration of the double sweep. (b) Zoom to observe the 4 data points acquired at a same potential ('down1', 'down2', 'up1' and 'up2'). Example from T104 at 966 km.

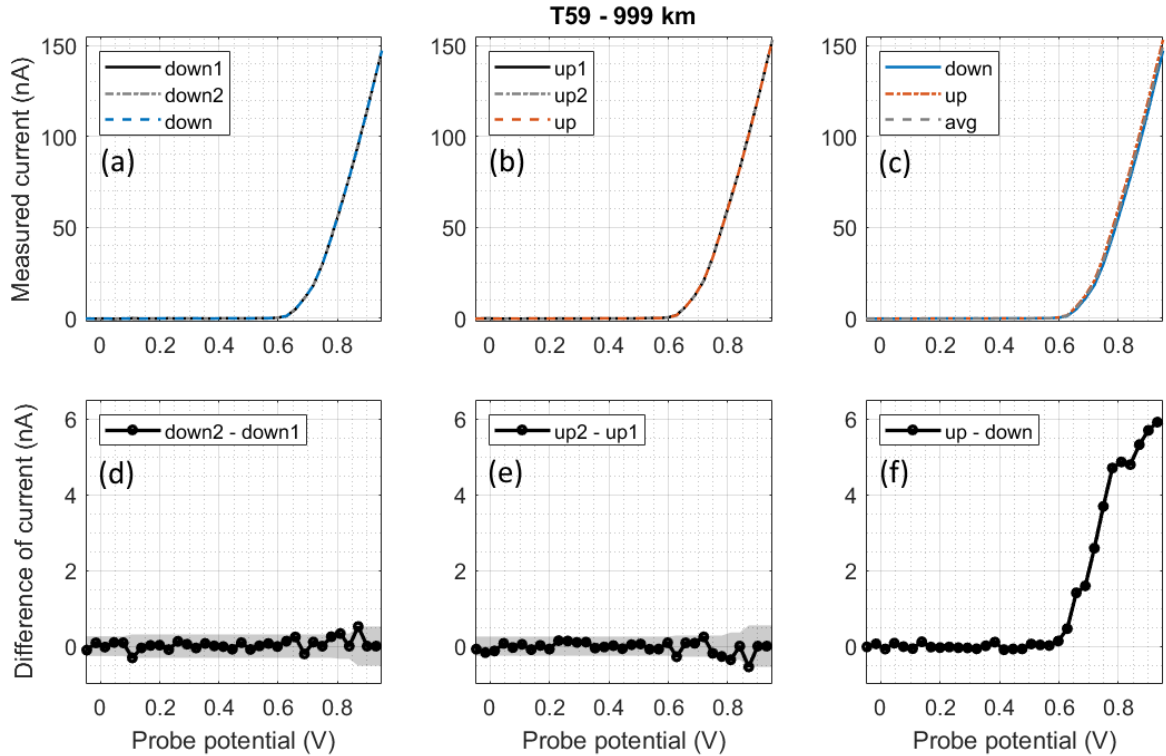

**Figure A2.** Raw data of the acquisition of a LP sweep. Left: data from the decreasing phase: (a) first point acquired ('down1'), second point acquired ('down2'), the average of both ('down') and (d) their difference ('down2 – down1'). The data error bar (in grey area) is deduced from this difference. Middle: the same for the increasing phase, respectively (b) and (e). Right: (c) the average for the decreasing ('down') and increasing ('up') phases, the average of both ('avg') and (f) their difference ('up – down'). From a measurement at ~999 km, during the Titan flyby T59.

The two data points taken at a same voltage step during the decreasing (respectively increasing) voltage phase 'down1' and 'down2' (respectively 'up1' and 'up2') are always very close to each other (inferior to 1% difference): the current stabilizes quickly at after each voltage step. The averaged curves 'down' and 'up' are globally identical (Figure A2c). It confirms that no organic surface coating is blurring the measurements. In the case of Figure A2, in the selected voltage range $I = [\text{Up1} - 0.7, \text{Up1} + 0.3]$ V (Up1 being the plasma potential of the first population, located at the beginning of the increase of the current), a maximum difference of current is observed at 6 nA, which represents ~4% of the total current (at ~150 nA).

Figure A3a shows the maximum values of 'down2 – down1', 'up2 – up1' and 'up – down' with altitude for 26 flybys on dayside, 22 flybys close to the terminator and 8 flybys on nightside far from the terminator. For each sweep, we selected the largest difference of current and the maximum of the averaged current in the voltage range $I = [\text{Up1} - 0.7, \text{Up1} + 0.3]$ V. Figure A3b shows their ratio (difference / average) as a function of altitude. In all cases, the difference 'down2 – down1' (blue triangles) and 'up2 – up1' (orange squares) are very small ($< 5$ nA, $< 2\%$ of the maximum current in $I$). The difference 'up – down' (black dots) is often positive. It is stronger in absolute value on dayside at lower altitude, but it is mainly because the electron current in these regions is higher. Compared to the maximum current intensity, the difference 'up – down' is inferior to 10% (with an average at 5%) in all altitude and SZA conditions. The stability of its value around 5% let us think that this small shift is due to a small electronic hysteresis. Nevertheless, this effect does not affect the results of this paper as the fitting done on 'down' and 'up' sweeps give the same values for the electron densities and temperatures (see example in Figure S1 in Supporting Information).

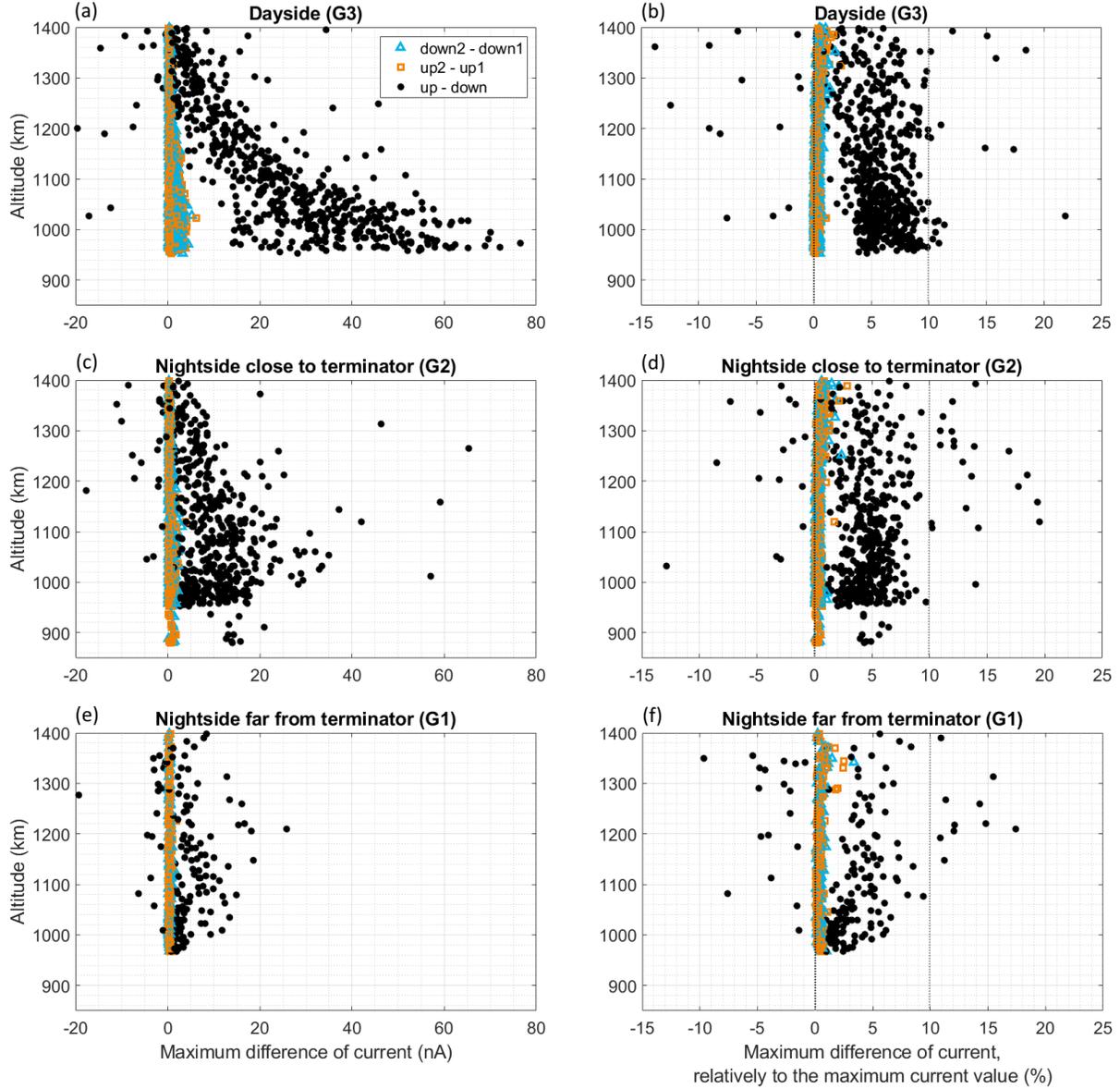

**Figure A3.** Maximum differences of the current ('down2 – down1', 'up2 – up1' and 'up – down') in the voltage range $I = [Up1 – 0.7, Up1 + 0.3]$ as a function of altitude (a,c,e). In subfigures (b,d,f), the maximum differences are divided by the maximum of the averaged current to facilitate the comparison at different altitudes (the 0% and 10% positions are highlighted with a thicker dotted line). (a,b) Flybys on dayside. (c,d) Flybys close to the terminator. (e,f) Flybys on nightside far from the terminator.

## Appendix B. Algorithm to fit the voltage sweeps with several populations

*B.1 Description of the fitting procedure*

To constrain the fitting of voltage sweeps in the case of several electron populations, the following automatic procedure is used. It is illustrated by an example in Figure B1.

① detection of the increase of the current (above 0.8 nA on dayside and above 0.4 nA on nightside) and of the saturated part ($U_{lim}$ being put 0.1 V after the maximum of the first derivative of the current);

② search for peaks in the second derivative of the current ($d^2I/dU^2$) in the voltage interval defined in ①;

③ selection of the part of the voltage sweep ($I_1$) with a voltage inferior to the potential corresponding to the minimum of $d^2I/dU^2$ between the two first peaks (or $< U_{lim}$ if there is only 1 population);

④ fit of the current curve for the first population with Equations (2) and (5) in the reduced voltage interval $I_1$;

If there is more than 1 population:

⑤ selection of the part of the curve ($I_2$) with a voltage inferior to the potential corresponding to the minimum of $d^2I/dU^2$ between peak 2 and peak 3 (or $< U_{lim}$ if there are only 2 populations);

⑥ fit of populations 1 and 2 with two set of Equations (2) and (5) in the reduced voltage interval $I_2$, the initial parameters for the fit of the first population being given by ④;

If there are more than 2 populations:

⑦ selection of the part of the curve ($I_3$) with a voltage inferior to the potential corresponding to the minimum of $d^2I/dU^2$ between peak 3 and peak 4 (or $< U_{lim}$ if there are only 3 populations);

⑧ fit of populations 2 and 3 with two set of Equations (2) and (5) in the reduced voltage interval $I_3$, the initial parameters for the fit of population 2 being given by ⑥. The contribution of population 1 is beforehand removed, based on the fit results in ⑥;

If there are more than 3 populations:

⑨ selection of the part of the curve ($I_4$) with a voltage inferior to $U_{lim}$;

⑩ fit of populations 3 and 4 with two set of Equations (2) and (5) in the reduced voltage interval $I_4$, the initial parameters for the fit of population 3 being given by ⑧. The contributions of populations 1 and 2 are beforehand removed, based on the fit results in respectively ⑥ and ⑧;

The fitting parts are performed with a trust-region-reflective least squares algorithm.

Only a few complex cases required a manual correction. It involves cases with two populations having close potentials $U_p$. These cases are discussed in detail in Section 3.2.3.2 and examples are given in Supporting Information Figures S8 and S9.

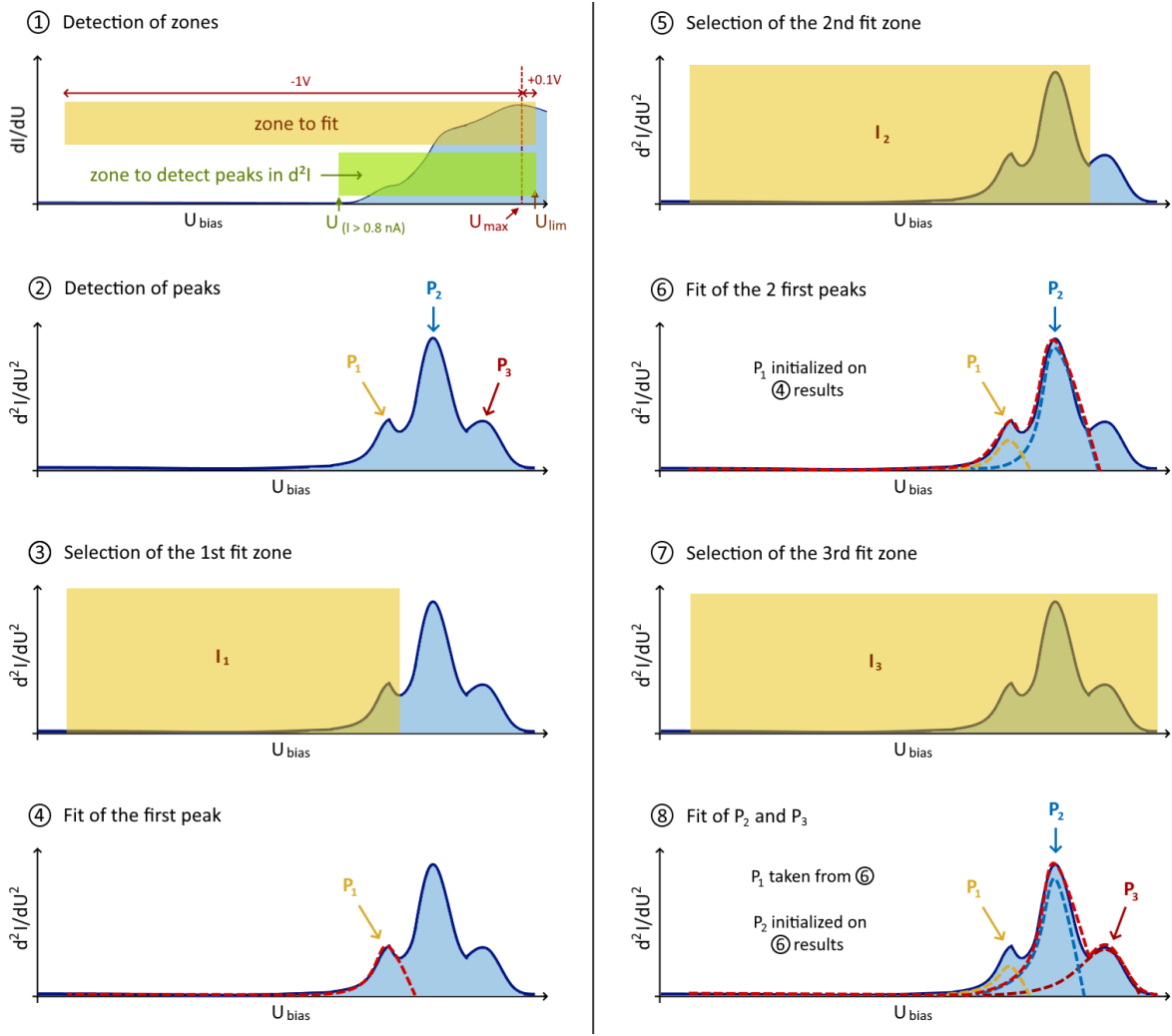

**Figure B1.** Illustration of the fitting procedure in the case of 3 populations. Note that we always fit the curve I(Ubias), not its derivatives. Derivatives for the plots are computed afterwards.

### B.2 Justification of the chosen algorithm

The algorithm presented above uses several fitting steps to help the fitting of smaller populations (eg. $P_1$ when $P_3$ is present). To validate this method, we compared its results to another algorithm (named 'MCMC') that uses only one fitting step. This second algorithm is based on Markov chain Monte Carlo simulations. It has two main advantages: (1) it simultaneously fits all the populations, and (2) it avoids local minima thanks to a Bayesian approach. The downside is that the MCMC algorithm takes 100 times longer to converge (eg. it requires 500 000 simulations to give an acceptable fit in the case of 3 populations).

Fitting results obtained with the two algorithms are very similar (see a representative example in Table B1). While the MCMC algorithm certainly does give lower uncertainties for the first small population, it does not significantly improve the rest of the results while demanding x100 more computational power and is not deemed necessary for the purpose of this study.

**Table B1.** Comparison of fit results using two different algorithms in the case of the sweep presented in Figures 2b and 3b (flyby T50, at 1125 km). Confidence intervals are given at 95%.

| Populations | Fit results | Algorithm used in this paper | Algorithm using Markov chain Monte Carlo simulations |
|---|---|---|---|
| $P_1$ | $T_e$ [meV] | $10 \pm 72$ | $16 \pm 6$ |
| | $n_e$ [cm$^{-3}$] | $24 \pm 269$ | $50 \pm 36$ |
| | $U_P$ [V] | $0.75 \pm 0.08$ | $0.757 \pm 0.006$ |
| $P_2$ | $T_e$ [meV] | $54 \pm 4$ | $45 \pm 7$ |
| | $n_e$ [cm$^{-3}$] | $498 \pm 75$ | $456 \pm 125$ |
| | $U_P$ [V] | $0.906 \pm 0.004$ | $0.904 \pm 0.004$ |
| $P_3$ | $T_e$ [meV] | $99 \pm 8$ | $96 \pm 10$ |
| | $n_e$ [cm$^{-3}$] | $433 \pm 69$ | $426 \pm 83$ |
| | $U_P$ [V] | $1.047 \pm 0.007$ | $1.047 \pm 0.004$ |

## Appendix C. Error bars

### C.1 Data error bars

Data error bars on the total current are estimated from the difference between 'down1' and 'down2' (or 'up1' and 'up2') measurements (see Figure A2). They increase with the probe voltage, with values similar to what was theorically predicted before the mission (Gurnett et al., 2004): ~100 pA for voltages inferior to Up1, and ~1 nA for voltages higher that Up1, with a smooth transition around Up1. These error bars do not vary with altitude, while the magnitude of the current generally increases towards lower altitudes. As a result, error bars are proportionally smaller at lower altitudes. This effect can be seen by comparing the data error bars on Figure 1, Figure 2 and Figure S3 (in Supporting Information), showing sweeps taken during T50 at different altitudes (respectively 1241 km, 1125 km and 1360 km).

The uncertainty on the fit of the ion current (confidence interval at 95%) is afterwards added to the error bars on the total current to obtain the error bars on the electron current. The three of them are plotted on Figure S3 in Supporting Information. The uncertainty on the ion current is at maximum 0.05 nA. It is always smaller than the total current error bars, and it is even negligible in most cases. This is due to the fact that the ion current is largely inferior to the electron current.

The fitting of the data is done on the electron current. However, as we plotted also its derivatives in the above figures for visual purposes, we also computed the error bars on dI/dU and d²I/d²U.

On a voltage sweep, at a voltage step U$_s$, the first derivative of the current is computed by:

$$\left(\frac{dI}{dU}\right)_s = \frac{I_{s+1} - I_{s-1}}{U_{s+1} - U_{s-1}} \qquad \text{(C-1)}$$

We used the propagation of uncertainties to deduce the error bars on the first derivative of the current $\sigma(dI/dU)$ at each voltage step:

$$\sigma\left(\frac{dI}{dU}\right)_s = \frac{2.\sigma(I)_s}{|U_{s+1} - U_{s-1}|} + 2.\sigma(U).\frac{|I_{s+1} - I_{s-1}|}{(U_{s+1} - U_{s-1})^2} \tag{C-2}$$

with $\sigma(I)_s$ being the electron current uncertainty at the voltage step $U_s$, and $\sigma(U)$ being the voltage uncertainty, equals to $5 \ 10^{-5}$ V. We observed that the second term of the equation (with $\sigma(U)$) is always negligible compared to the first one (with $\sigma(I)$).

We used the same technique for the second derivative:

$$\frac{d^2I}{dU^2}\bigg)_s = \frac{dI/dU)_{s+1} - dI/dU)_{s-1}}{U_{s+1} - U_{s-1}} \tag{C-3}$$

$$\sigma\left(\frac{d^2I}{dU^2}\right)_s = \frac{2.\sigma(dI/dU)_s}{|U_{s+1} - U_{s-1}|} + 2.\sigma(U).\frac{|dI/dU)_{s+1} - dI/dU)_{s-1}|}{(U_{s+1} - U_{s-1})^2} \tag{C-4}$$

### C.2 Confidence intervals for fit results

The parameters fitted are not directly $(U_P, n_e, T_e)$, but $(U_P, I_0, \beta = 1/T_e)$, with $T_e$ the electron temperature in electron volt. We developed a matlab code for the fit. From the computed Jacobian we deduced the confidence intervals at 95% for $(U_P, I_0, \beta)$. The standard deviation is then computed for $n_e$ and $T_e$ by propagation of the uncertainties. Concerning $n_e$, the standard deviation $(\sigma_{n_e})$ is derived from Equation (3) applied to electrons.

$$\sigma_{n_e} = \left(\frac{\sqrt{2\pi m_e}}{A_{LP}e^{3/2}}\right) \times \left(\sigma_{I_0}.\sqrt{\beta} + I_0.\frac{\sigma_\beta}{2\sqrt{\beta}}\right) \tag{C-5}$$

$$\sigma_{T_e} = \frac{\sigma_\beta}{\beta^2} \tag{C-6}$$

where $\sigma_{I_0}$ and $\sigma_\beta$ are the standard deviations of respectively $I_0$ and $\beta$ (obtained by the fit).

## Appendix D. Link between the second derivative of the electron current and the Electron Energy Distribution Function (EEDF)

This section explains the link between the second derivative of the current (d²I/dU²) and the Electron Energy Distribution Function (EEDF). It aims to justify the attribution of electrons populations to peaks in the second derivative. The presented method finally gives the Maxwellian EEDF deduced from the fitting of each of the electron populations detected on a same measurement.

### D.1 The Druyvesteyn method

The Druyvesteyn method (Druyvesteyn, 1930; Lieberman & Lichtenberg, 2005) was used to compute the EEDF from the electron current measured by the Langmuir probe. The method is valid for any convex probe geometry and does not depend on the probe dimension compared to Debye length, or on the ratio $T_i/T_e$.

Here we detail how to obtain the equation linking the EEDF to the second derivative of the current. The equation for the electron velocity distribution as function of the second derivative of the electron current is given by the equation (3) in Druyvesteyn (1930):

$$\mathcal{V}\,(\Delta U) = \frac{4\,m_e}{A_{LP}\,e^2}\cdot \Delta U \cdot \frac{d^2 I_e}{d\Delta U^2} \qquad\qquad [(m.s^{-1})^{-1}.m^{-3}] \qquad\qquad \text{(D-1)}$$

Electrons with a speed between $v$ and $v + dv$ have energies between $E$ and $E + dE$, therefore:

$$\mathcal{V}(v).dv = EEDF(E).dE \qquad\qquad \text{(D-2)}$$

The derivative of the speed with respect to energy is given by:

$$E = \frac{1}{2}m_e v^2 \quad => \quad \frac{dv}{dE} = (2m_e E)^{-\frac{1}{2}} \qquad\qquad \text{(D-3)}$$

Therefore, we obtain the EEDF by dividing the electron speed distribution by $\sqrt{2m_e E}$, with $E(\Delta U) = -e\,.\,\Delta U$. The EEDF is thus computed from the second derivative of the current, and the electron density is equal to the integral of the EEDF:

$$EEDF_{Dr}\,(\Delta U) = \frac{2\sqrt{2}}{A\,e^2}\sqrt{\frac{m\,.(-\Delta U)}{e}}\,\frac{d^2 i}{d\Delta U^2} \qquad [J^{-1}.m^{-3}]$$

$$with\;\; n_e = \int_0^{+\infty} EEDF_{Dr}\big(E(\Delta U)\big)\,.dE \qquad\qquad \text{(D-4)}$$

If the Druyvesteyn method is applied to the expression of the current given in Equation (2), it gives back the expression of an EEDF for a Maxwellian speed distribution:

$$EEDF_{Maxw}(E) = \frac{2}{\sqrt{\pi}}.(k_B.T_e)^{-\frac{3}{2}}.n_e.\sqrt{E}.\exp\left(-\frac{E}{k_B.T_e}\right) \qquad\qquad \text{(D-5)}$$

### D.2 EEDF: example with one population

This section presents a simple case with one electron population (introduced in Figure S3 and S5(a) in Supporting Information). Figure D1 plots the EEDF obtained by applying the Druyvesteyn method to the data points (Equation (D-4)), and the 'Maxwellian' EEDF computed with the electron density and temperature resulting from the fit (Equation (D-5)). They do not perfectly match (the maximum of the peak is shifted). This is because of the low resolution of the data points. Indeed, the Druyvesteyn method uses the second derivative of the current, which is highly distorted because of the low number of data points (e.g. the sharp initial increase is lowered by this effect).

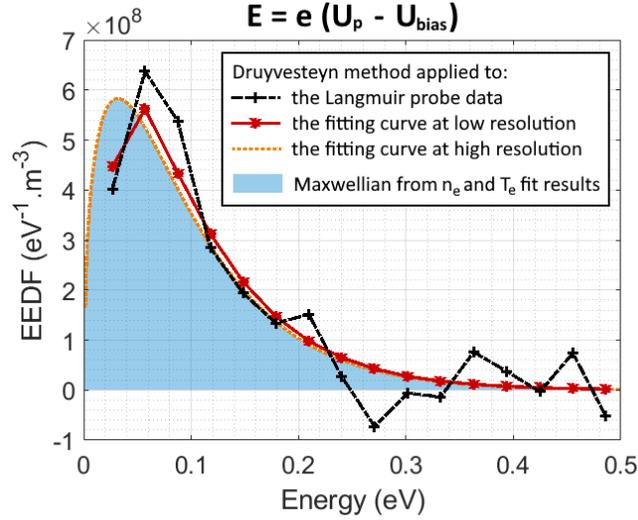

**Figure D1.** Example of EEDF obtained from Langmuir probe data with only one population using the Druyvesteyn method (dotted-dashed black line). This curve is compared to the EEDF obtained with the Druyvesteyn method applied to fitting curves at low resolution (red line) and at high resolution (orange dotted line). Finally, there is also the 'Maxwellian' EEDF computed with $n_e$ and $T_e$ resulting from the fit (blue area). To compare to Figure S5a in Supporting Information that gives the plot of d²I/dU² in the same case. Data acquired during the inbound of T50 at 1360 km.

To show the effect of resolution, the Druyvesteyn method is applied to the fit of the current, from which we can extract artificial datasets at a chosen numerical resolution. The EEDF obtained using a fitting curve at high resolution (> 200 points per peak) superimposes very well with the 'Maxwellian' EEDF. On the other side, the EEDF obtained using a fitting curve with the resolution of the Langmuir probe data points (4-10 points per peak) matches well with the EEDF obtained from the data points. The EEDF obtained using the Druyvesteyn method on the Langmuir probe data points is therefore fairly well represented by a 'Maxwellian' EEDF if one takes the problem of low resolution into account.

### D.3 EEDF: example with several populations

In the case of data showing several populations, applying the Druyvesteyn method is less straightforward: the method requires one 'plasma potential' $U_P$, whereas the fit gives as much $U_P$ as populations detected. Figures D2 and D3 show the case of two populations, respectively with an explicative scheme and real data. The Druyvesteyn method is applied to the data points using the $U_P$ potential from (a) the first population and (b) the second population. In Figures D2② and D3b, the two population peaks appear in reverse order compared to the plot of d²I/dU², respectively in Figure D2① and Figure 3a, because the x-axis is reversed ($E = -e \cdot \Delta U \propto -U_{bias} + U_P$).

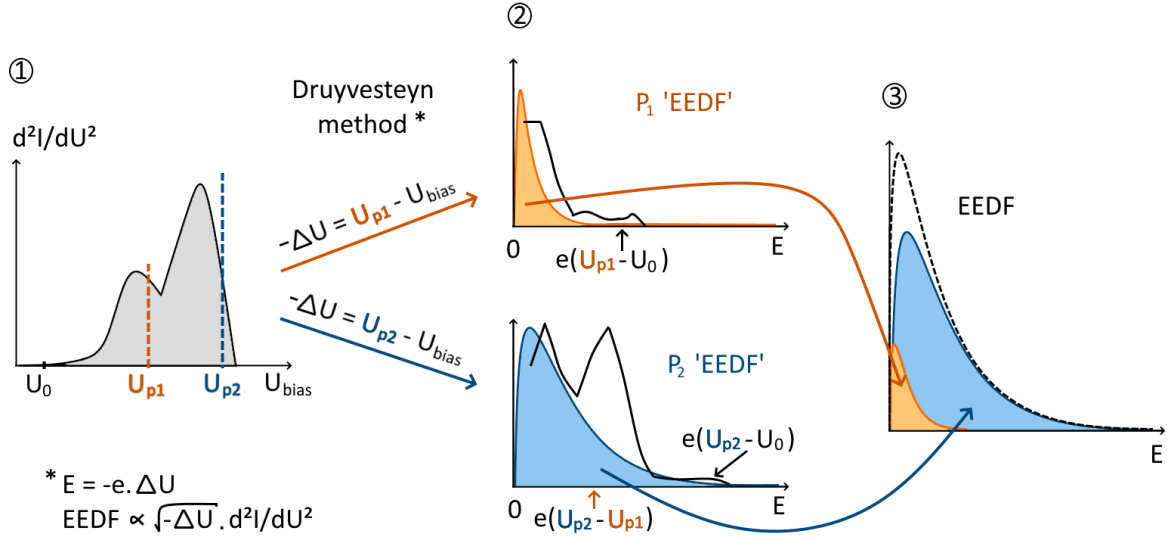

**Figure D2.** Explanation of how to obtain Figure D3 and Figure D4. ① We start from the I(U$_{bias}$) data, fitted to obtain $(U_p, n_e, T_e)$ for each population (see Figure 3). ② The Druyvesteyn method is applied to the I(U$_{bias}$) data (black line) and to the fitting curve for each population (colored areas), similarly to Figure D3. Note that the x-axis is reversed compared to ①. ③ 'EEDF' obtained for each population in ② are compared and added to give the total EEDF, as in Figure D4.

Figure D3 shows that in the case of several electron populations, the EEDF computed for one population with the Druyvesteyn method is disturbed by the other populations with close values of $U_P$. To observe this effect, the Druyvesteyn method is applied to the total fitted current (orange line), and to the fitted current corresponding only to one population (dashed blue line). The difference between these two curves is the effect of nearby populations, observed on both Figures D3a and D3b. Therefore, if one takes into account both the presence of nearby populations (supplementary peak at higher energy) and the effects of low resolution (shift of the main peak), the EEDF from data points can reasonably be modelled by a 'Maxwellian' EEDF (shaded area). An example with 3 populations is given in Supporting Information Figure S6.

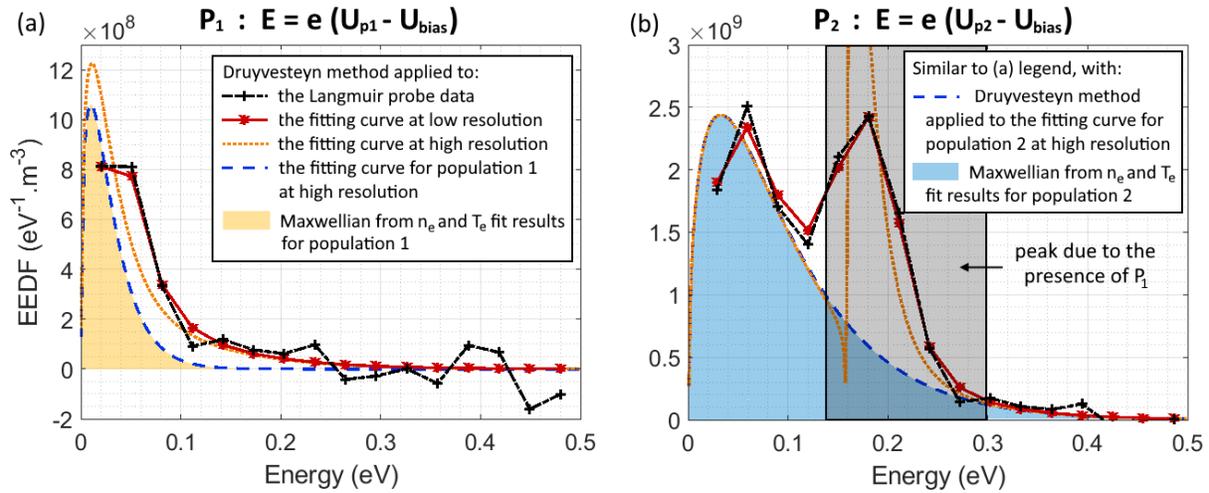

**Figure D3.** Example of EEDF obtained from Langmuir probe data with two populations using the Druyvesteyn method: (a) for population 1; (b) for population 2. Comparison of the EEDF using the Druyvesteyn method on data points (dotted-dashed black line) to the EEDF obtained with the Druyvesteyn method applied to the fitting curves at low resolution (red line), at high resolution (orange dotted line), and to the fitting curve of only one population at high resolution (dashed blue line). Finally, there are also the 'Maxwellian' EEDF computed with $n_e$ and $T_e$ for each population resulting from the fit (yellow and blue areas). Figure to compare to Figure 3a that gives the plot of d²I/dU² in the same case. Data acquired during the inbound of T50 at 1241 km.

### D.4 Diversity of EEDF combinations for the different populations

The 'Maxwellian' EEDF of the electron populations are compared in Figure D4. As mentioned in Section D.1, the area under each EEDF corresponds to the electron density of each population and the position of the maximum of the EEDF is related to the electron temperature.

For clarity, Figure D4 only shows fit results, corresponding to the part ③ of Figure D2. Its objective is to show the variability of the electron distribution (i.e. the changes in electron populations) for different flybys. It illustrates three very different cases obtained at ~1000 km altitude, which are representative of the three flyby groups (G1, G2 and G3) presented in details in Section 3.1. The first case (from the group G1) shows two populations (named $P_1$ and $P_2$). The area (and so the electron density) of $P_2$ dominates $P_1$. The second case (G2) has one additional population ($P_3$). The area of $P_2$ and $P_3$ are equivalent and higher than the area of $P_1$. We observe that $P_3$ distribution is in average at higher energy than the distribution of $P_2$: its electron temperature is higher. Finally, the group G3 shows still another population ($P_4$) and $P_1$ becomes negligible compared to the others. $P_3$ dominates $P_2$ in density. $P_4$ characteristics vary with altitude: it can be denser than the others, and it is usually hotter.

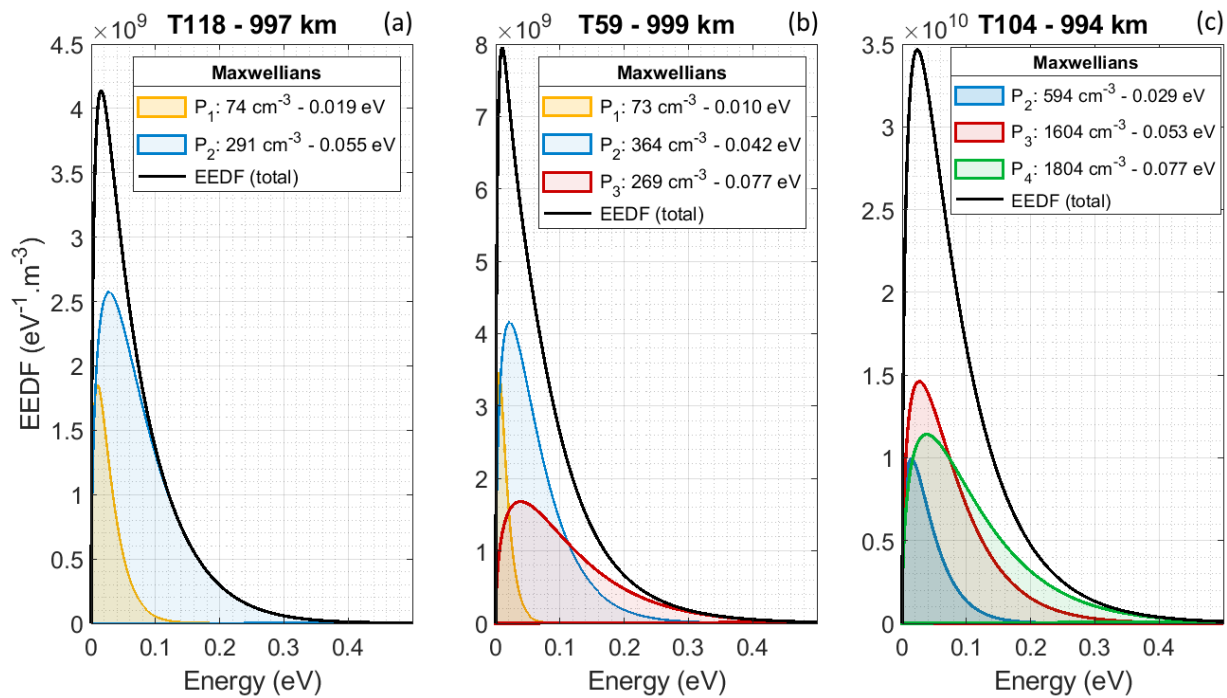

**Figure D4.** EEDF obtained from Maxwellian populations, at ~1000 km altitude for the flybys: (a) T118 (group G1); (b) T59 (G2); (c) T104 (G3).